\documentclass[11pt]{article}
\usepackage{amssymb}
\usepackage{epsfig}

\newcommand{\BE}{\begin{equation}}
\newcommand{\EE}{\end{equation}}
\newcommand{\BA}{\begin{eqnarray}}
\newcommand{\EA}{\end{eqnarray}}

\begin{document}
\begin{titlepage}

\vspace*{1mm}
\begin{center}

   {\LARGE{\bf Precision test for the new Michelson-Morley  \\
experiments with rotating cryogenic cavities }}

\vspace*{14mm}
{\Large  M. Consoli}
\vspace*{4mm}\\
{\large
Istituto Nazionale di Fisica Nucleare, Sezione di Catania \\
Via Santa Sofia 64, 95123 Catania, Italy}
\end{center}
\begin{center}
{\bf Abstract}
\end{center}
A new ether-drift experiment in D\"usseldorf is currently measuring
the relative frequency shift of two cryogenic optical 
resonators upon active rotations of the apparatus. I point out that
the observed fractional 
amplitude of the sidereal variations of the signal in February, 
$C_{\rm sid}\sim (11 \pm 2) \cdot 10^{-16}$, is entirely consistent 
with the expectations based on Miller's observations in the same
epoch of the year. This leads to
predict that, with future data collected in August-September, the observed
sidereal variations should increase by $\sim + 70 \%$, i.e. up to 
$C_{\rm sid}\sim (19\pm 2)\cdot 10^{-16}$
retaining the present normalization. This would represent clean experimental 
evidence for the existence of a preferred frame. 
 
\end{titlepage}


{\bf 1.}~Modern ether-drift experiments 
look for a preferred frame by
measuring the relative frequency shift $\delta \nu$ between two 
cavity-stabilized lasers upon local rotations \cite{brillet}
or under the Earth's rotation \cite{muller}. Today, by combining the 
possibility of active rotations of the apparatus 
with the use of cryogenic optical resonators, a new experiment \cite{schiller}
is currently trying to push the relative accuracy of the measurements to the
level ${\cal O}(10^{-16})$. It is generally believed that such an extremely 
high precision, that strongly constrains the anisotropy of the speed of 
light, can also rule out the idea of a preferred frame.

However, as discussed in Refs.\cite{pla,cimento} 
(see also \cite{pagano}), 
vanishingly small values of $\delta \nu$, by themselves, 
cannot rule out the old Lorentzian formulation
of Relativity. In fact, if light propagates isotropically in a preferred frame
$\Sigma$, the relative frequency shift observed on the Earth is given by 
\BE
{{\delta \nu}\over{ \nu}} 
\sim 
({\cal N}_{\rm medium} -1) 
{{v^2_{\rm earth}}\over{c^2}}
\EE
where
${\cal N}_{\rm medium}\sim 1$ is the refractive index of the medium where light 
propagation takes place and
$v_{\rm earth}$ denotes the Earth's velocity with respect to $\Sigma$
(value projected in the plane of the interferometer).  

This frequency shift
vanishes identically when the speed of light coincides with the same parameter
"c" entering Lorentz transformations, i.e. in an ideal vacuum where 
${\cal N}_{\rm vacuum}=1$. Starting from this observation, 
one can re-read \cite{pla,cimento} the classical 
and modern ether-drift experiments in a consistent framework where 
there is a preferred frame relatively to which the Earth is moving with a
velocity $\sim 200$ km/s (value projected in the plane of the interferometer). 
This velocity is consistent with the values deduced by Miller
(see Table V of Ref.\cite{miller}) on the base of the sidereal
variations of the ether-drift effect in different periods of the year. 

The aim of this Letter is to show that the sidereal variations 
observed in February by the authors of
Ref.\cite{schiller} are completely consistent with Miller's
observations in the same epoch of the year. 
This also suggests a precise prediction to be tested with future data collected
in the period of August-September. A confirmation of this prediction would
represent clean experimental evidence for the existence of a preferred frame.
\vskip 15 pt
{\bf 2.}
The basic ingredient for our analysis is the relation \cite{pla}
for the frequency shift of two orthogonal optical resonators
 to ${\cal O}({{v^2_{\rm earth}}\over{c^2}})$ 
\BE
\label{basic}
              {{\delta \nu (\theta) }\over{\nu}}= 
{{\bar{u}'(\pi/2 +\theta)- \bar{u}' (\theta)} \over{u}} \sim 
|B_{\rm medium}| {{v^2_{\rm earth} }\over{c^2}} \cos(2\theta)
\EE
Here $\theta=0$ indicates the direction of the ether-drift,  
$\bar{u}'(\theta)$ is
the two-way speed of light (as measured on the Earth)
in a medium of refractive index ${\cal N}_{\rm medium}\sim 1$, 
$B_{\rm medium}\sim -3({\cal N}_{\rm medium} -1)$ and 
$u=c/{\cal N}_{\rm medium}$
indicates the isotropical value. Eq.(\ref{basic}) is 
derived under the assumption that light is seen isotropical in 
a preferred frame $\Sigma$ and that the required trasformation law 
to the Earth's frame is a Lorentz transformation. 

In this way one obtains an amplitude of the ether-drift effect
\BE
\label{amplitude}
{{A_{\rm th}(t)}\over{\nu}}=|B_{\rm medium}| {{v^2_{\rm earth}(t) }\over{c^2}}
\EE
that vanishes identically in a vacuum.

However, as originally suggested in Ref.\cite{pagano}, even
in the case of an extremely high vacuum (such as the one existing in the
resonating cavities of Ref.\cite{schiller}) 
 one can argue that the
speed of light differs from the parameter "c" entering Lorentz transformations.
In fact, for an apparatus
placed on the Earth's surface (but otherwise
in free fall with respect to any other
gravitational field), General Relativity predicts a tiny refractive index
\BE
\label{nphi}
            {\cal N}_{\rm vacuum}\sim  1- 2\varphi
\EE
with 
\BE 
     \varphi =- {{G_N M_{\rm earth}}\over{c^2 R_{\rm earth} }} \sim
-0.7\cdot 10^{-9}
\EE
so that
\BE
\label{bvacuum}
                 B_{\rm vacuum} \sim 6\varphi\sim  -4.2 \cdot 10^{-9}
\EE
In this way, for a mean reference value $v_{\rm earth}=210$ km/s, and a 
mean
frequency $\nu \sim 2.8\cdot 10^{14}$ Hz, as in Ref.\cite{schiller}, 
Eq.(\ref{amplitude}) implies
\BE
\label{reference}
\langle A_{\rm th} \rangle \sim  0.58~{\rm Hz}
\EE
Let us now compare this with the basic Eq.(1) of
Ref.\cite{schiller} for the
frequency shift at a given time $t$ 
\BE
\label{basic2}
    {{\delta \nu [\theta(t)]}\over{\nu}} = 
\hat{B}(t)\sin 2\theta(t) + 
\hat{C}(t)\cos 2\theta(t)
\EE
where $\theta(t)$ is the angle of rotation of the apparatus, 
 $\hat{B}(t)\equiv 2B(t)$ and $\hat{C}(t)\equiv 2C(t)$ so that
one finds an experimental amplitude
\BE
\label{expampli}
A_{\rm exp}(t)= \nu \sqrt { \hat{B}^2(t) + \hat{C}^2(t) }
\EE
Taking into account the experimental results \cite{schiller}
$\langle \hat{B}\nu \rangle \sim 2.8$ Hz and
$\langle \hat{C}\nu \rangle \sim -3.3$ Hz, one finds
\BE
\label{expampli2}
\langle A_{\rm exp}\rangle \sim 4.3~~{\rm Hz}
\EE
which is much larger than the theoretical prediction in Eq.(\ref{reference}).
Therefore, as suggested by the same authors of Ref.\cite{schiller}, 
for a meaningful comparison, one has to subtract the mean value and restrict
to the sidereal modulations of the signal. 

To predict the variations of the ether-drift effect, I shall
use the average
data and the theoretical curves reported by Miller in Figs.26 of
Ref.\cite{miller}. To compare with the data collected around February 6th 
(of 2005) by the authors of Ref.\cite{schiller}, I'll also
restrict to Miller's observations around February 8th (of 1926). In this
case, the sidereal variations of the ether-drift effect are
modest and the profile of the observable velocity is found in the range
7.5 km/s $\leq v_{\rm obs}(t) \leq$ 10 km/s. Using Eq.(13) of 
Ref.\cite{pla}, $v^2_{\rm obs} \sim 3({\cal N}^2_{\rm medium}-1) v^2_{\rm earth}$  
(for Miller's interferometer that was operating in air where 
${\cal N}_{\rm air}\sim 1.00029$), this
range of {\it observable} velocities is found to correspond to the range of 
{\it kinematical} velocities 
\BE
180~{\rm km/s}\leq v_{\rm earth}(t) \leq 240 ~{\rm km/s}
\EE
Therefore, taking into account Eqs.(\ref{amplitude}) and (\ref{bvacuum}), the
amplitude of the signal should lie in the range
0.42 Hz$\leq A_{\rm th}(t)\leq$ 0.76 Hz, with an 
overall daily variation
\BE
\label{daily}
\Delta A_{\rm th} \sim \pm 0.17~{\rm Hz}
\EE
that represents a fractional change 
\BE
\label{frac}
 {{\Delta A_{\rm th} }\over{\nu}}\sim \pm 6 \cdot 10^{-16}
\EE
An experimental check of this prediction can be found in Table
I of Ref.\cite{schiller}. There one finds a clear indication for a non-zero
value of the parameter 
\BE
C_{\rm sid}\equiv \sqrt{C^2_1 + C^2_2}
\EE
 that controls the sidereal modulation of the signal
\BE
 \hat{C}(t)=C_0 + 
C_1\sin(\omega_{\rm sid}t)+ C_2\cos(\omega_{\rm sid}t)+...
\EE
In this case, the result
\BE
          C_{\rm sid} \sim (11 \pm 2) \cdot 10^{-16}
\EE
implies a daily variation of the amplitude
\BE
\Delta A_{\rm exp} \sim \pm \nu C_{\rm sid} \sim \pm 0.31~{\rm Hz}
\EE 
that is entirely consistent with the theoretical value in Eq.(\ref{daily}). 

By itself, this result might be considered a first 
experimental check of the ether-drift observations reported by Miller. However, 
for a real precision test, one has still to
wait a few months. In fact, looking
again at Figs.26 of Ref.\cite{miller}, one discovers that the sidereal 
variations in August-September were 
 considerably larger with a profile
of the observable velocities lying now in the range
5.0 km/s $\leq v_{\rm obs}(t) \leq$ 10 km/s. This range corresponds to
\BE
120~{\rm km/s}\leq v_{\rm earth}(t) \leq 240 ~{\rm km/s}
\EE
and to
\BE
\Delta A_{\rm th} \sim \pm 0.29~{\rm Hz}
\EE
with fractional changes of $\sim \pm 10\cdot 10^{-16}$. In this case, the
fitted value of the coefficient $C_{\rm sid}$ should increase by $\sim +70 \%$. 
Retaining the present normalization, this means up to
\BE
          C_{\rm sid}\sim (19 \pm 2) \cdot 10^{-16}
\EE
that represents a 3-4 $\sigma$ change relatively to the present value and
might be observable with future data taken in the next few months. 
\vskip 15 pt
{\bf 3.}
The new generation of Michelson-Morley experiments with
 rotating cryogenic optical resonators will likely definitely resolve, in one 
way or the other, the 
old controversy (Einsteinian vs. Lorentzian) on
the interpretation of the relativistic effects. In this Letter, 
I have illustrated the implications of Miller's observations 
for the present
experiment in D\"usseldorf of Ref.\cite{schiller} and the forthcoming one
in Berlin \cite{peters}. The present amplitude of the
sidereal variations observed in February 2005 by the authors of 
Ref.\cite{schiller}, as embodied in the parameter 
$C_{\rm sid}\sim (11\pm 2)\cdot 10^{-16}$, is completely consistent 
with the prediction based on Miller's data collected in the
same epoch of the year. 

However, for a real definitive check, the
fitted value of $C_{\rm sid}$ should also increase by $\sim + 70\%$ 
when taking data in August-September, i.e. up to 
$C_{\rm sid}\sim (19 \pm 2)\cdot 10^{-16}$ retaining the same normalization. 
This prediction will be tested in the
next few months and, whenever confirmed, would represent clean experimental 
evidence for the existence of a preferred frame. If this will happen, 
further checks
will require to determine the azimuth of the ether-drift effect and, with it, 
the cosmic component 
of the Earth's velocity. To this end, the data should be analyzed in a 
model-independent way without necessarily restricting the preferred frame
to coincide with the CMBR.

\vfill
\eject

\end{document}